\documentclass[twocolumn,showpacs,superscriptaddress,floatfix]{revtex4}
\usepackage{graphicx}%
\usepackage{dcolumn}
\usepackage{amsmath}

\makeatletter
\def\btt#1{\texttt{\@backslashchar#1}}%
\DeclareRobustCommand\bblash{\btt{\@backslashchar}}%
\makeatother

\begin{document}

%\preprint{HEP/123-qed}

\title{Charge-based silicon quantum computer architectures \\using controlled single-ion implantation}% Force line breaks with \\

\author{A.~S.~Dzurak}\email{a.dzurak@unsw.edu.au}\affiliation{Centre for Quantum Computer Technology}\affiliation{Schools of Physics
and Electrical Engineering, University of New South Wales, NSW
2052, Australia}

\author{L.~C.~L.~Hollenberg}\affiliation{Centre for Quantum Computer Technology}\affiliation{School of
Physics, University of Melbourne, VIC 3010, Australia}

\author{D.~N.~Jamieson}\affiliation{Centre for Quantum Computer Technology}\affiliation{School of
Physics, University of Melbourne, VIC 3010, Australia}

\author{F.~E.~Stanley}\affiliation{Centre for Quantum Computer Technology}\affiliation{Schools of Physics
and Electrical Engineering, University of New South Wales, NSW
2052, Australia}

\author{C.~Yang}\affiliation{Centre for Quantum Computer Technology}\affiliation{School of
Physics, University of Melbourne, VIC 3010, Australia}

\author{T.~M.~B\"{u}hler}\affiliation{Centre for Quantum Computer Technology}\affiliation{Schools of Physics
and Electrical Engineering, University of New South Wales, NSW
2052, Australia}

\author{V.~Chan}\affiliation{Centre for Quantum Computer Technology}\affiliation{Schools of Physics
and Electrical Engineering, University of New South Wales, NSW
2052, Australia}

\author{D.~J.~Reilly}\affiliation{Centre for Quantum Computer Technology}\affiliation{Schools of Physics
and Electrical Engineering, University of New South Wales, NSW
2052, Australia}

\author{C.~Wellard}\affiliation{Centre for Quantum Computer Technology}\affiliation{School of
Physics, University of Melbourne, VIC 3010, Australia}

\author{A.~R.~Hamilton}\affiliation{Centre for Quantum Computer Technology}\affiliation{Schools of Physics
and Electrical Engineering, University of New South Wales, NSW
2052, Australia}

\author{C.~I.~Pakes}\affiliation{Centre for Quantum Computer Technology}\affiliation{School of
Physics, University of Melbourne, VIC 3010, Australia}

\author{A.~G.~Ferguson}\affiliation{Centre for Quantum Computer Technology}\affiliation{Schools of Physics
and Electrical Engineering, University of New South Wales, NSW
2052, Australia}

\author{E.~Gauja}\affiliation{Centre for Quantum Computer Technology}\affiliation{Schools of Physics
and Electrical Engineering, University of New South Wales, NSW
2052, Australia}

\author{S.~Prawer}\affiliation{Centre for Quantum Computer Technology}\affiliation{School of
Physics, University of Melbourne, VIC 3010, Australia}

\author{G.~J.~Milburn}\affiliation{Centre for Quantum Computer Technology}\affiliation{Department of Physics, University of Queensland, QLD 4072,
Australia}

\author{R.~G.~Clark}\affiliation{Centre for Quantum Computer Technology}\affiliation{Schools of Physics
and Electrical Engineering, University of New South Wales, NSW
2052, Australia}

%\author{A. S. Dzurak${\rm ^{1*}}$, L. C. L. Hollenberg${\rm ^{2}}$, D. N. Jamieson${\rm ^{2}}$, F. E. Stanley${\rm
%^{1}}$, C. Yang${\rm ^{2}}$, T. M. B\"{u}hler${\rm ^{1}}$, V.
%Chan${\rm ^{1}}$, D. J. Reilly${\rm ^{1}}$, C. Wellard${\rm
%^{2}}$, A. R. Hamilton${\rm ^{1}}$, C. I. Pakes${\rm ^{2}}$, A. G.
%Ferguson${\rm ^{1}}$, E. Gauja${\rm ^{1}}$, S. Prawer${\rm ^{2}}$,
%G. J. Milburn${\rm ^{3}}$ and R. G. Clark${\rm ^{1}}$}

%\affiliation{Centre for Quantum Computer Technology ${\rm
%^{1}}$Schools of Physics and Electrical Engineering, University of
%New South Wales, NSW 2052, Australia ${\rm ^{2}}$School of
%Physics, University of Melbourne, VIC 3010, Australia ${\rm
%^{3}}$Department of Physics, University of Queensland, QLD 4072,
%Australia}\email{a.dzurak@unsw.edu.au}

%\author{A. S. Dzurak${\rm ^{*}}$, F. E. Stanley, T. M. B\"{u}hler, V. Chan, D. J. Reilly, A. R. Hamilton, A. G. Ferguson, E. Gauja, R. G. Clark}
%\affiliation{Centre for Quantum Computer Technology, Schools of
%Physics and Electrical Engineering, University of New South Wales,
%NSW 2052, Australia}\email{a.dzurak@unsw.edu.au}

%\author{L. C. L. Hollenberg, D. N. Jamieson, C. Yang, C. Wellard, C. I. Pakes, S. Prawer}
%\affiliation{Centre for Quantum Computer Technology, School of
%Physics, University of Melbourne, VIC 3010, Australia}

%\author{G. J. Milburn}
%\affiliation{Centre for Quantum Computer Technology, Department of
%Physics, University of Queensland, QLD 4072, Australia}

\date{\today}% It is always \today, today, but you may specify any date with \date.

\begin{abstract}
We report a nanofabrication, control and measurement scheme for
{\it charge}-based silicon quantum computing which utilises a new
technique of {\it controlled} single ion implantation. Each qubit
consists of two phosphorus dopant atoms ${\rm \sim}$50 nm apart,
one of which is singly ionized. The lowest two energy states of
the remaining electron form the logical states. Surface electrodes
control the qubit using voltage pulses and dual single electron
transistors operating near the quantum limit provide fast readout
with spurious signal rejection. A low energy (keV) ion beam is
used to implant the phosphorus atoms in high-purity Si. Single
atom control during the implantation is achieved by monitoring
on-chip detector electrodes, integrated within the device
structure, while positional accuracy is provided by a nanomachined
resist mask. We describe a construction process for implanted
single atom and atom cluster devices with all components
registered to better than 20 nm, together with electrical
characterisation of the readout circuitry. We also discuss
universal one- and two-qubit gate operations for this
architecture, providing a possible path towards quantum computing
in silicon.
\end{abstract}


\maketitle

\section{INTRODUCTION}
It is widely believed that solid-state systems have much to offer
in the search for a scalable quantum computer (QC) technology. One
of the most advanced proposals is based on superconducting qubits
\cite{a1}, where coherent control \cite{a2,a3,a4,a5,a6} and
two-qubit coupling \cite{a7} have been demonstrated. Semiconductor
schemes also show promise, in particular those based on silicon
metal-oxide-semiconductor (Si MOS) technology, due to their
compatibility with existing manufacturing. In Kane's scheme
\cite{a8} (Fig. 1(a)) the qubits are defined by long-lived nuclear
spin states of buried  phosphorus dopants in a Si host crystal,
and manipulated by external surface gates and radio-frequency (RF)
magnetic fields. An alternative Si:P architecture has also been
proposed using electron spin states as qubits \cite{a9}. While
measurement of single nuclear and electron spins in the
solid-state has very recently been reported for diamond
\cite{a10}, an electronic method of single spin readout in silicon
remains a significant challenge.

Although detection of single spins in silicon has not been
achieved, fast single {\it charge} detection is already in place
due to recent developments in RF single-electron transistor (SET)
technology \cite{a11,a12}. We have therefore proposed a Si:P
charge qubit architecture \cite{a13} which is complementary to the
Kane scheme, but experimentally accessible now. Charge based
qubits in large lithographically-defined quantum dots have been
proposed previously \cite{a14,a15}. In our system the quantum
logic states correspond to the lowest two states of the single
valence electron localised in the double well formed by two donor
phosphorus atoms (see Fig. 1(b)).

\begin{figure}[h]
%\vspace{1cm}
\centering
\includegraphics[width=8.5cm]{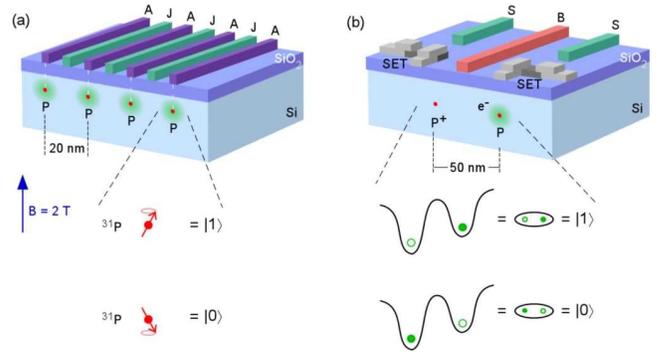}
\caption{Quantum computer architectures based on buried dopants in
silicon. (a) Kane scheme \cite{a8}, where a qubit is defined by
the spin of a single ${^{31}}$P nucleus. (b) Charge-based scheme
\cite{a13}, discussed here, where the position of a single
electron within a double-well potential created by {\it two} P
dopants defines logical states ${\rm \mid}$0${\rm \rangle}$ and
${\rm \mid}$1${\rm \rangle}$. Dual single electron transistors
(SETs) on the surface enable qubit readout.}
\end{figure}\label{fig1}

\pagebreak

Gates above the buried ${\rm P-P^{+}}$ system allow for external
control over the barrier height ({\it B}-gate) and potential
symmetry ({\it S}-gates) for manipulation of localised qubit
states, $ {\rm {\mid}0{\rangle} = {\mid}L{\rangle}} $ and ${\rm
{\mid}1{\rangle}  = {\mid}R{\rangle}}$, while twin RF SETs
facilitate qubit initialization and readout. A thin barrier layer
such as SiO$ {\rm _{2}}$ is used to isolate the gates from the
donors below. The charge qubit will decohere faster than its
spin-based counterpart due to stronger environmental coupling,
however, calculated \cite{a13} gate operation times of order 50 ps
are commensurately faster than the ${\mu}$s operation times
\cite{a16} for spin qubits. Although not spin-dependent, the
electron transfer process in the ${\rm P-P^{+}}$ qubit is similar
to that in the ${\rm P-P}$ system used for spin readout in the
Kane scheme \cite{a8} and as such provides a test bed for the
future development of spin-based qubits.

Operational issues aside, the spin- and charge-based Si:P schemes
shown in Fig. 1 face an identical challenge -- placement of
individual phosphorus donors within a low-disorder
intrinsic-silicon ($i$-Si) substrate at precise array sites,
accurately positioned with respect to the surface control gates
and SETs. Two contrasting approaches are being developed to attain
this goal. In the first, scanned-probe lithography of a
hydrogenated silicon surface together with epitaxial Si overgrowth
may be used to construct a buried P array using a bottom-up atomic
assembly approach \cite{a17,a18}. While potentially capable of
atomic precision, a number of steps remain in the full
demonstration of this technology.

Here we describe a second construction approach for which charge
qubit test devices have been fabricated. Array sites are defined
by top-down lithographic patterning of a resist mask and the
donors are implanted through this using a keV phosphorus ion beam,
with on-chip single ion detection. We have introduced the concept
of single ion implantation in previous work on alternate device
structures \cite{a18a} and here show the first experimental
demonstration of the technique for 14 keV phosphorus ions, as
required for Si:P qubit construction.

The article is organised as follows. We first describe the process
by which the implantation of P donors is controlled with single
ion accuracy, allowing devices to be configured atom-by-atom.
Ion-impact detectors, integrated into the devices, are monitored
electrically during implantation. When a single ion enters the
$i$-Si substrate it produces electron-hole pairs that drift in an
applied electric field, creating a detectable current pulse for
each ion strike. Although discussed here for construction of QC
devices, this technology is applicable to any semiconductor device
where accurate control of dopant number in the substrate is
important. We then detail a complete fabrication process for the
construction of test devices with donors implanted with precision
beneath control gates and SETs and describe the use of
cross-correlated RF twin-SET measurements which will be necessary
for qubit readout. Finally, we outline a scheme for the operation
of one- and two-qubit devices accessible using this technology.

\section{Controlled single-ion implantation}
Fig.~1(a) depicts the technique developed to localize individual
phosphorus atoms at the desired qubit array sites. A
nano-patterned ion-stopping resist such as polymethyl-methacrylate
(PMMA) defines the array sites and a low-energy keV ${\rm
^{31}P^{+}}$ ion beam is used to implant the P dopants through the
thin SiO$ {\rm _{2}}$ barrier layer to a mean depth of 20 nm into
the substrate.

\begin{figure}[h]
\centering
\includegraphics[width=8.5cm]{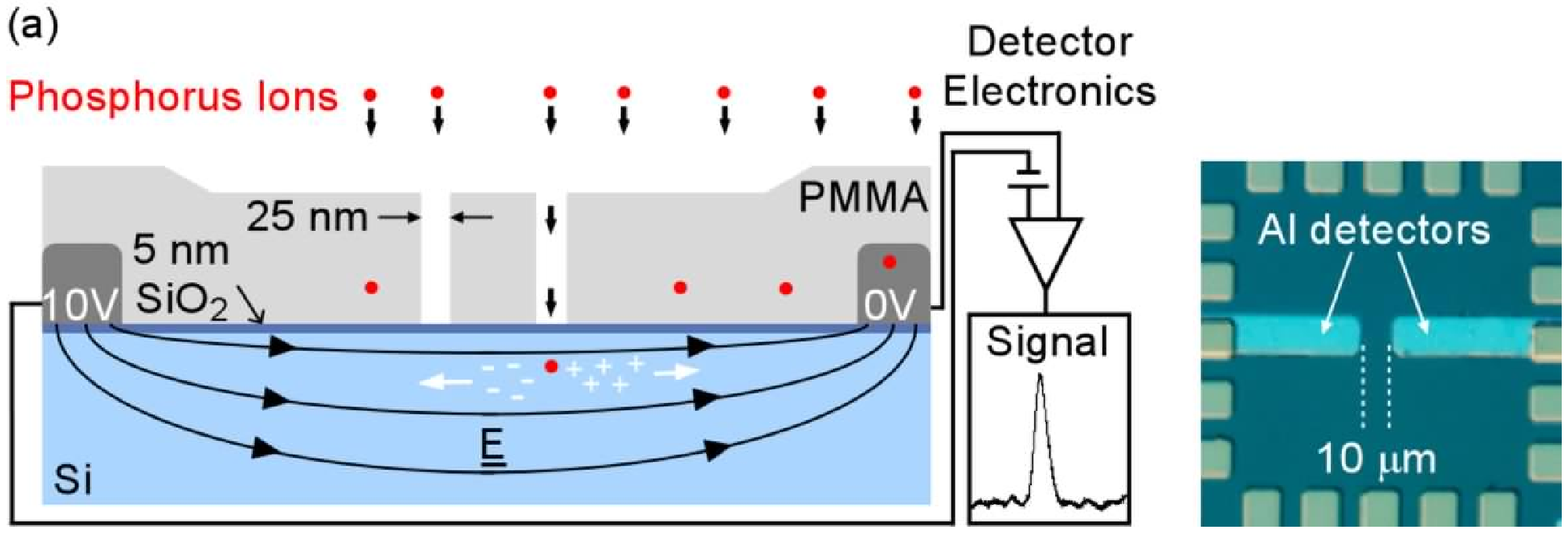}\\
\includegraphics[width=8.5cm]{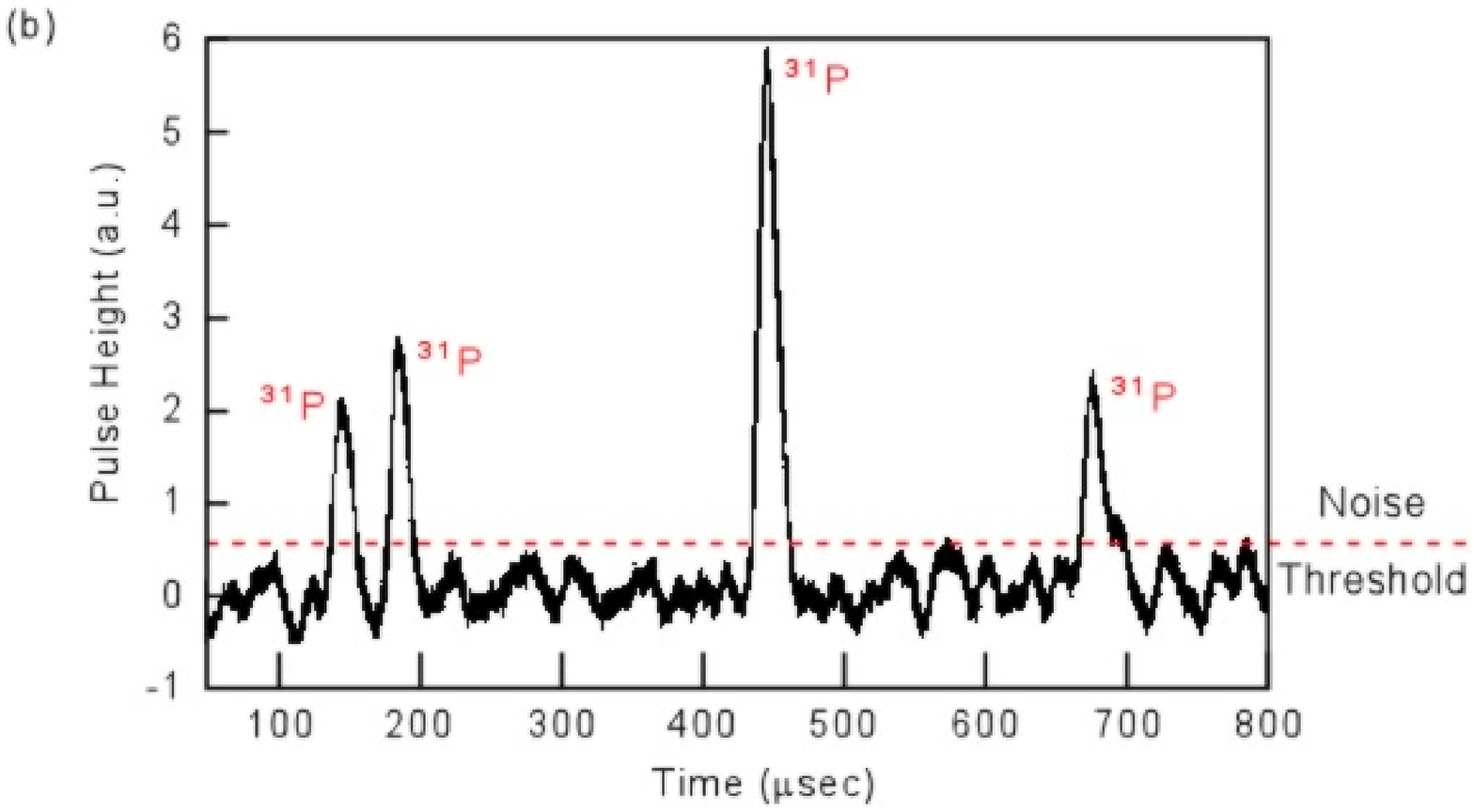}\\
\includegraphics[width=8.5cm]{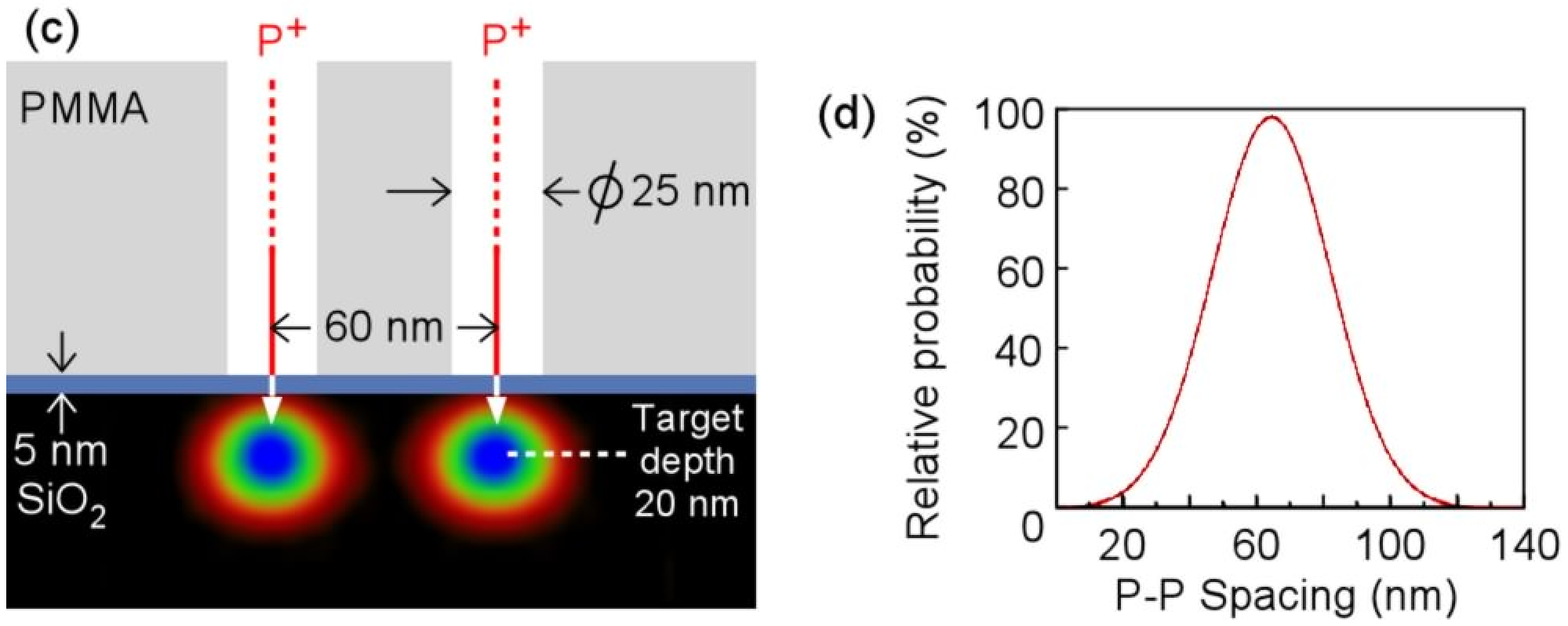}
\caption{(a) Schematic (not to scale) showing phosphorus
implantation through a resist mask. Each ion strike creates an
electron-hole plasma, producing a current pulse monitored by
on-chip electrodes (shown in optical micrograph at right). (b)
Experimental demonstration of single ion strikes in an $i$-Si
substrate bombarded with a 14 keV phosphorus ion beam. \\ (c)
Calculated 14 keV P${\rm ^{+}}$ implant position probability for
two 25 nm apertures separated by 60 nm. (d) Resulting histogram of
spacings between pairs of P${\rm ^{+}}$ implants through the
apertures.}
\end{figure}\label{fig2}

Single ion impacts are detected by two surface aluminium detector
electrodes, biased at up to 10 V, which form MOS tunnel junctions
and act as back-to-back diodes, restricting the dark current (no
${\rm ^{31}P^{+}}$ beam) to below 100 pA. Each ion entering the
substrate creates ${\rm \sim}$500 e$^{-}$/h$^{+}$ pairs which
drift in the internal electric field produced by the electrode
bias. The e$^{-}$/h$^{+}$ pair generation and separation mechanism
has been modelled \cite{a19} using the semiconductor modeling
packages SRIM \cite{a20} and TCAD \cite{a21} and is found to
create a current transient in 40 ps. The current is integrated in
an external, high efficiency cooled preamplifier circuit to
produce a single pulse for each ion strike. The substrate is also
cooled to 115 K to reduce detector noise. Since the collection
time is much less than the recombination time, close to 100\%
charge collection is possible. To measure the efficiency of the
detectors, we have rastered a focused MeV ion beam across various
electrode geometries and monitored the charge collection
efficiency at each point \cite{a22}, finding charge collection
efficiencies of ${\rm \sim}$99\% at distances up to 10
${\mathrm{\mu}}$m from the electrodes. We have therefore been able
to fabricate detector electrodes set back many microns from the
central nanostructured region where the qubits and control gates
are located (see image in Fig. 2(a)).

Data for an incident 14 keV ${\rm ^{31}P^{+}}$ ion beam is shown
in Fig. 2(b) for an interdigitated electrode array with lateral
dimension ${\rm \sim}$100 ${\rm \mu}$m. Pulses above the noise
threshold occur at a frequency consistent with the areal ion dose
of ${\rm \sim}$5 ions/ms across the active device area and can
therefore be identified as single ion strikes, as confirmed by
analysis of the pulse height distribution, discussed below. The
temporal broadening of the peaks in Fig. 2(b) results from the
time constant of the detection circuit.

While semiconductor detectors are well-established \cite{a23}, to
our knowledge this is the first use of a detector integrated into
a device to control its doping level. Furthermore, the 14 keV data
of Fig. 2(b) represents the first detection of implanted keV ${\rm
^{31}P^{+}}$ ions, although light ions such as ${\rm ^{1}H^{+}}$
and ${\rm ^{4}He^{+}}$ have been previously measured in the range
15-30 keV using silicon detectors \cite{a24}. The low noise and
high energy resolution of our detectors results from their low
leakage current together with the very thin SiO$ {\rm _{2}}$
dead-layer, confirmed to be less than 7 nm from measurements using
MeV ions \cite{a22}.

Because of their mass, ${\rm ^{31}P^{+}}$ ions with incident
energy 14 keV lose ${\rm \sim}$85\% of their initial energy to
nuclear stopping events in the substrate. The energy available for
the creation of e$^{-}$/h$^{+}$ pairs is therefore ${\rm \sim}$2
keV. Due to variations in ion trajectories the exact number of
e$^{-}$/h$^{+}$ pairs and the resulting pulse height will vary
between events, as observed experimentally in Fig. 2(b). Analysis
of a large number of pulses shows a gaussian distribution of peak
heights with a mean value consistent with the expected electronic
energy loss, confirming each peak as a single ion impact event. To
ensure that no ions enter the substrate without detection the
noise level must be kept significantly below the mean pulse
height. We have been able to reduce the effective noise level to
an energy equivalent of ${\rm \sim}$1 keV and expect that below
0.2 keV should be attainable, providing greater than 99\%
confidence that all implanted ions have been detected.

At present we are applying a uniform areal ion dose to our masked
substrates, so that ion placement is random between the apertures.
For the ${\rm P-P^{+}}$ charge qubit device of Fig. 1(b) this
leads to 50\% probability of correctly configuring a device with
one P atom at each site. Such a yield is sufficient for
proof-of-principle experiments on one-qubit devices, however, for
large-scale qubit arrays it will be necessary to direct each ion
to its appropriate array site using a focused ion beam (FIB). We
are currently developing this technology using a dual beam FIB/SEM
with a 20 nm focus.

Because the path taken through the substrate by each implanted ion
is different, there will be variation in the spatial configuration
of each ${\rm P-P^{+}}$ pair, which must be corrected for by
appropriate calibration of the $B$- and $S$-gate voltages (defined
above) for all qubits in the device. Since the qubit gate
operation times are dependent on both the ${\rm P-P}$ donor
spacing and the barrier gate voltage \cite{a13}, the $B$-gate can
be used to tune individual qubits over a wide range of spacings.

To estimate the spacings in our test devices we have used an
implant modeling package \cite{a20} to calculate the expected ion
straggle. For 14 keV ${\rm ^{31}P^{+}}$ ions incident on a silicon
substrate with a 5 nm SiO$ {\rm _{2}}$ gate oxide we find that the
ions come to rest 20 nm below the free surface, with a standard
deviation of 10 nm in the beam direction and 7 nm in the lateral
direction. We have also modeled a typical implant profile for a
device with two circular resist apertures of diameter 25 nm and a
centre-to-centre spacing of 60 nm. Fig. 2(c) shows the calculated
position probability contours, while Fig. 2(d) is the resulting
relative probability of ${\rm P-P}$ donor spacings. Our
calculations indicate that the magnitude of the detection pulse is
related to the donor depth, meaning that devices with two similar
pulses would be closer to optimum configuration and could then be
selected for measurement.

At the mean distance of 60 nm we calculate \cite{a13} qubit
rotation times of ${\rm \sim}$200 ps using accessible $B$-gate
voltages, becoming faster for smaller spacings. As seen in Fig.
2(d), for initial two-donor devices ${\rm \sim}$40\% of ${\rm
P-P}$ pairs will be separated by less than 60 nm, giving
functional qubits. We have fabricated resist apertures as small as
15 nm and expect that centre-to-centre spacings of 30 nm can be
achieved. In such structures all pair spacings would be below 70
nm, providing a high yield of operational qubits for large scale
systems.

\section{Fabrication of Test Devices}
To evaluate the potential of charge qubits constructed via ion
implantation we have fabricated test devices in which the two
donors in Fig. 1(b) are replaced by implanted {\it clusters} of
phosphorus donors, effectively creating two buried metallic
islands. Fig. 3 shows devices incorporating two such clusters
buried 20 nm below the surface, accurately aligned to control
gates and dual read-out SETs. We calculate that approximately 600
donors are required in each cluster to create a metallic density
of states separated by a barrier, enabling periodic sequential
tunneling {\it between} clusters upon application of a
differential bias between the surface control gates. The two
read-out SETs may then be used to detect this periodic charge
motion. The presence of a {\it periodic} output signal in test
devices would provide an unequivocal demonstration of controlled
electron transfer, while the large number of implanted donors
minimises the detrimental effects of any traps in the $i$-Si
substrate or at the Si/SiO${\rm _{2}}$ interface.

\begin{figure}
\includegraphics[width=8.5cm]{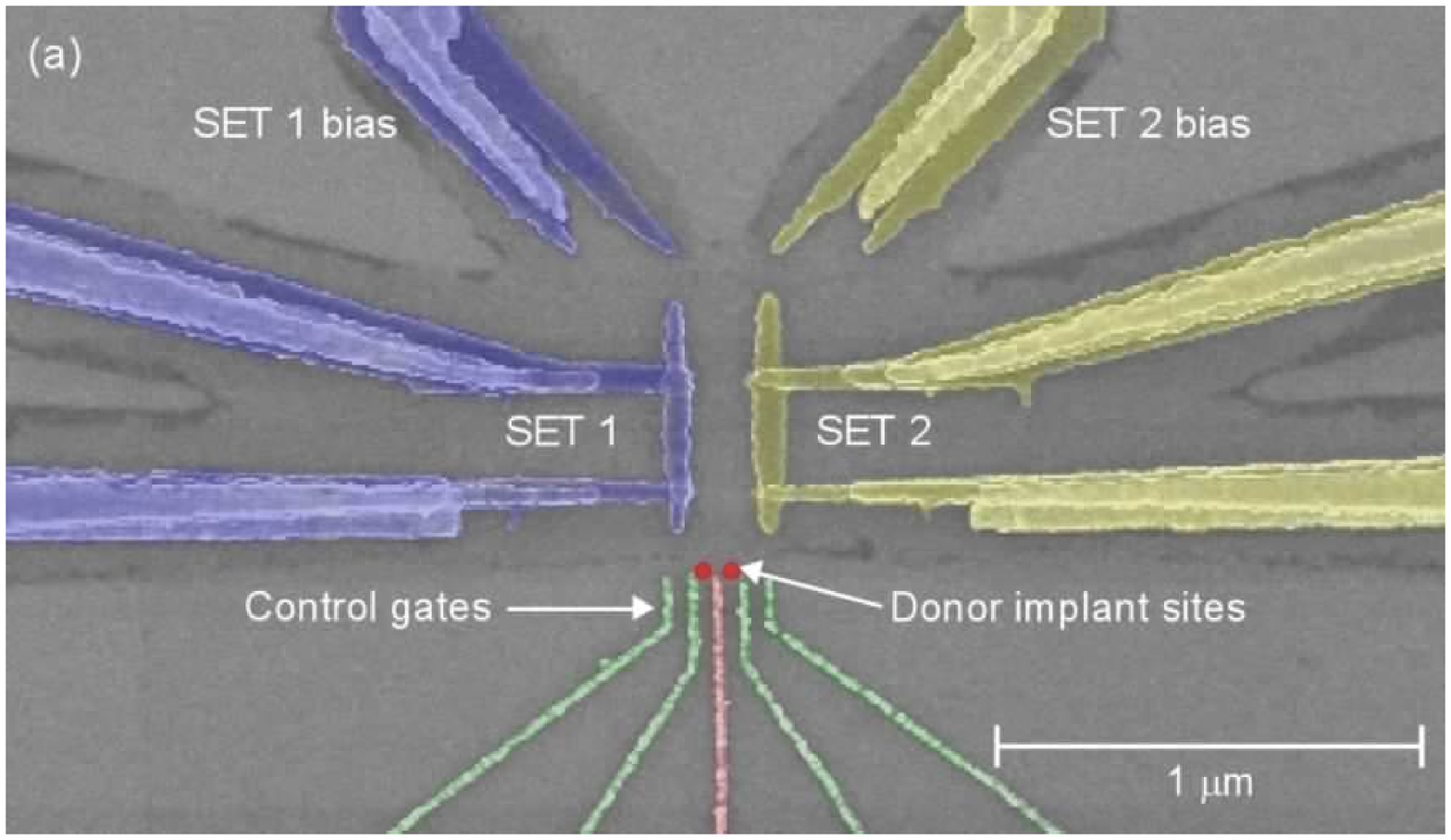}\\
\centering \includegraphics[width=8.5cm]{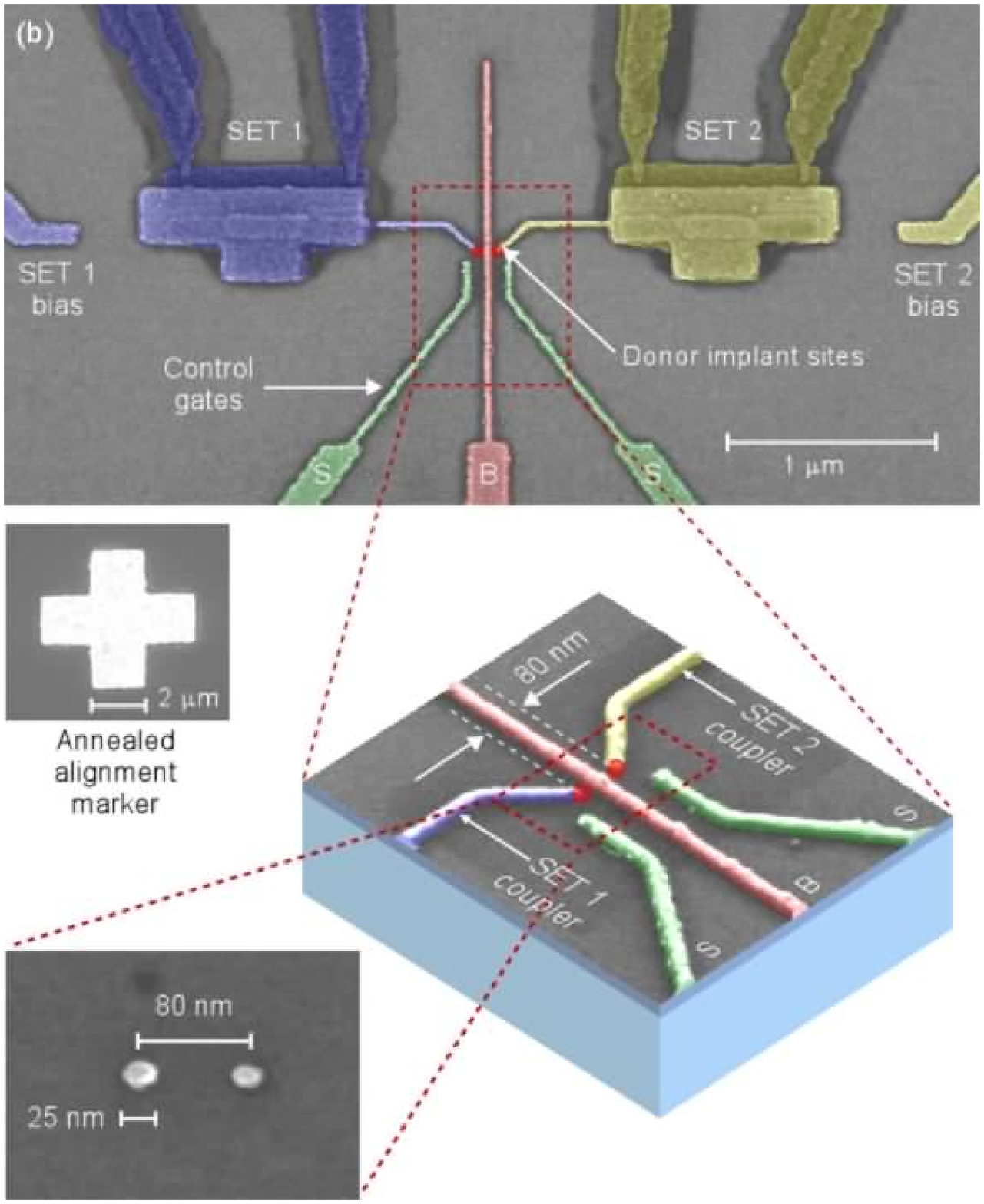}
\vspace{0.5cm}\caption{SEM images of test devices incorporating
buried phosphorus ion implants, shown {\it schematically} here by
red markers. (a) Device with five control gates (bottom of image)
to control electron motion between donors, together with two
closely-spaced SETs to detect electron transfer. (b) Modified
device, where each SET island has a coupling electrode to
capacitively couple it to a phosphorus donor site. A long central
barrier ($B$) gate may be used to control tunneling between
donors. {\it Bottom left:} Demonstration, via metal lift-off, of
25 nm PMMA implantation apertures. }
\end{figure}\label{fig3}

\addvspace{1cm}

Fabrication of the devices in Fig. 3 involves a number of
high-resolution electron beam lithography (EBL) steps, each of
which must be aligned to the others with an accuracy of 20 nm or
better, to ensure reliable gate control and sufficient capacitive
coupling between the donors and read-out SETs. The process flow
proceeds as follows. Firstly, a barrier between the control gates
and donors is provided by a 5nm SiO$ {\rm _{2}}$ layer, thermally
grown on a near-intrinsic silicon wafer with a background
$n$-doping level of ${\rm 10^{12} cm^{-3}}$.

If single ion implant control is required, micron-scale aluminium
detector electrodes are then deposited on the substrate using UV
lithography. The electrodes are separated from each other by 10
${\rm{\mu}}$m (see Fig. 2(a)), sufficiently close to ensure
high-efficiency charge collection, but far enough apart to allow
all nanocircuitry to be constructed between them. We note that for
the fabrication of test devices with large numbers of P atoms,
single ion counting was not required as the cluster size could be
determined sufficiently accurately from the incident ion flux and
resist-aperture diameter.

To provide sub-20 nm registration between all features on the
device, EBL is then used to pattern a number of Ti/Pt alignment
markers on the chip. In a second EBL step, two sub-30 nm apertures
are opened in an ion-stopping PMMA resist, as depicted
schematically in Fig. 2(a). Metallisation and lift-off results
shown in the inset to Fig. 3(b) confirm the dimensions of these
apertures.

Donor implantation proceeds next using a 14 keV ${\rm P^{+}}$ ion
beam. Modeling \cite{a20} indicates that these ions come to rest
in the PMMA at a mean depth of 38 nm, with standard deviation 10
nm, so a layer thickness above 100 nm is sufficient to block
phosphorus ions and avoid forward recoils of atoms constituting
the resist. Phosphorus ions which pass through the apertures and
enter the substrate come to rest at a mean depth of 20 nm below
the free surface, as stated above.

Damage to the substrate caused by the implant process must next be
removed via a thermal treatment. We employ a 950${\rm ^{\circ}}$C
rapid thermal anneal (RTA) for 5 seconds, sufficient to activate
the phosphorus donors \cite{a25} but limiting their diffusion to
${\rm \sim}$1 nm based on standard bulk rates \cite{a26}. Pulsed
laser annealing on ms timescales could also be used to further
limit phosphorus diffusion and to localise the region of heating.

Following ion implantation and activation, the remaining
nanocircuitry on the surface of the chip is completed using two
further EBL steps. Firstly, Ti/Au control gates are deposited
following EBL patterning of a single PMMA layer. We routinely
fabricate gate widths of 20-30 nm using this process \cite{a27}
and have also demonstrated continuous gates  as narrow as 12 nm.
Finally, the two Al/Al${\rm _{2}}$O${\rm _{3}}$ SETs are
fabricated using a double-angle metallisation process and a
bilayer resist \cite{a28}. As seen in Fig. 3(a), the overall
alignment between all levels of this process is better than the
width of a control gate (${\rm \sim}$20 nm).

\section{Twin-SET readout and electrical characterisation}
The proposed charge qubit device of Fig. 1(b) employs {\it two}
symmetric SETs to read out electron position within a single ${\rm
P-P^{+}}$ qubit. Whilst in principle a single SET would suffice,
by cross-correlating the output from two SETs it is possible to
reject spurious events resulting from random charge motion within
the Si substrate, the SiO${\rm _{2}}$ barrier layer, or associated
with the SETs themselves. Such {\it charge noise} rejection has
been demonstrated \cite{a29} using an all-aluminium twin-SET
architecture in which the phosphorus donors are simulated by two
aluminium islands separated by a tunnel junction. In the device of
Fig. 3(a) each SET is designed to capacitively couple most
strongly to its nearest donor cluster. Preliminary measurements,
however, indicated significant cross-coupling between SETs, making
signal discrimination difficult, and it was necessary to
reconfigure the device with the two SETs separated from each other
by around 1 ${\rm \mu}$m (Fig. 3(b)). Additional antenna
electrodes are then used to couple each SET to its target donor or
donor cluster, while a long central barrier ($B$) gate increases
screening between SETs. In the following we focus on
characterisation of the control and readout circuitry of this
modified device with no implanted donors.

To read out the state of a charge qubit it will be necessary to
perform a projective measurement with the two SETs on a timescale
shorter than the qubit mixing time \cite{a12} $t_{mix}$, which we
expect to be of order microseconds based on spontaneous emission
results obtained for superconducting circuits \cite{a30}. Using
radio-frequency (RF) SET technology \cite{a11,a12} we have
developed a twin RF-SET measurement system \cite{a31} and used it
to demonstrate cross-correlated single-shot measurements of
controlled charge motion within all-aluminium devices on
sub-microsecond timescales \cite{a32}. The time $\tau_{meas}$
required to perform single-shot measurements of electron position
on buried donor devices will depend on the capacitive coupling
between the donor electrons and SETs \cite{a29,a32}. Estimates of
this coupling \cite{a33} indicate that $\tau_{meas}$ ${\rm \sim}$1
${\rm \mu}$s can be achieved, making qubit readout possible.

\begin{figure}[t] \centering
\includegraphics[width=8.5cm]{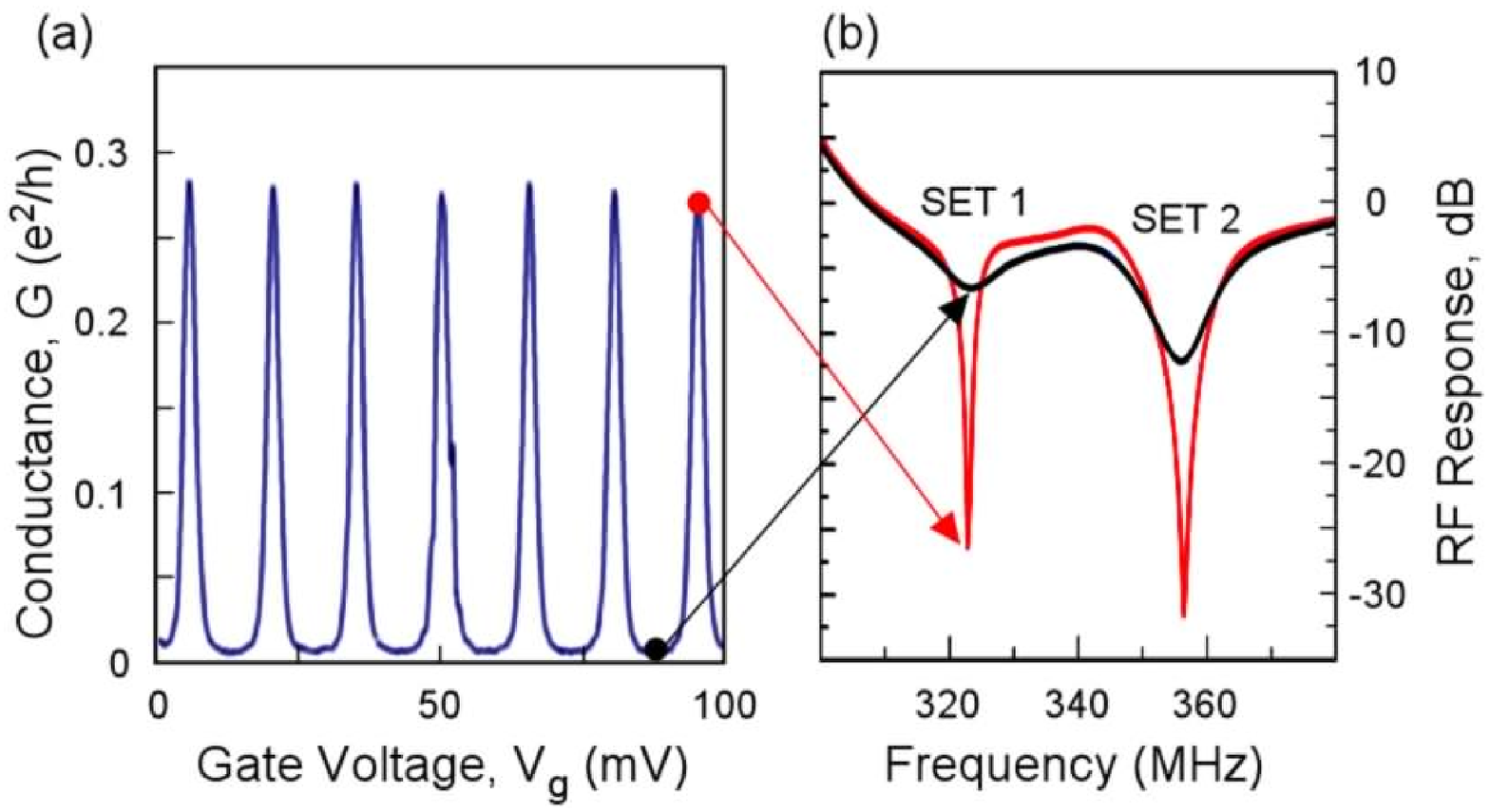}\\
\includegraphics[width=8.5cm]{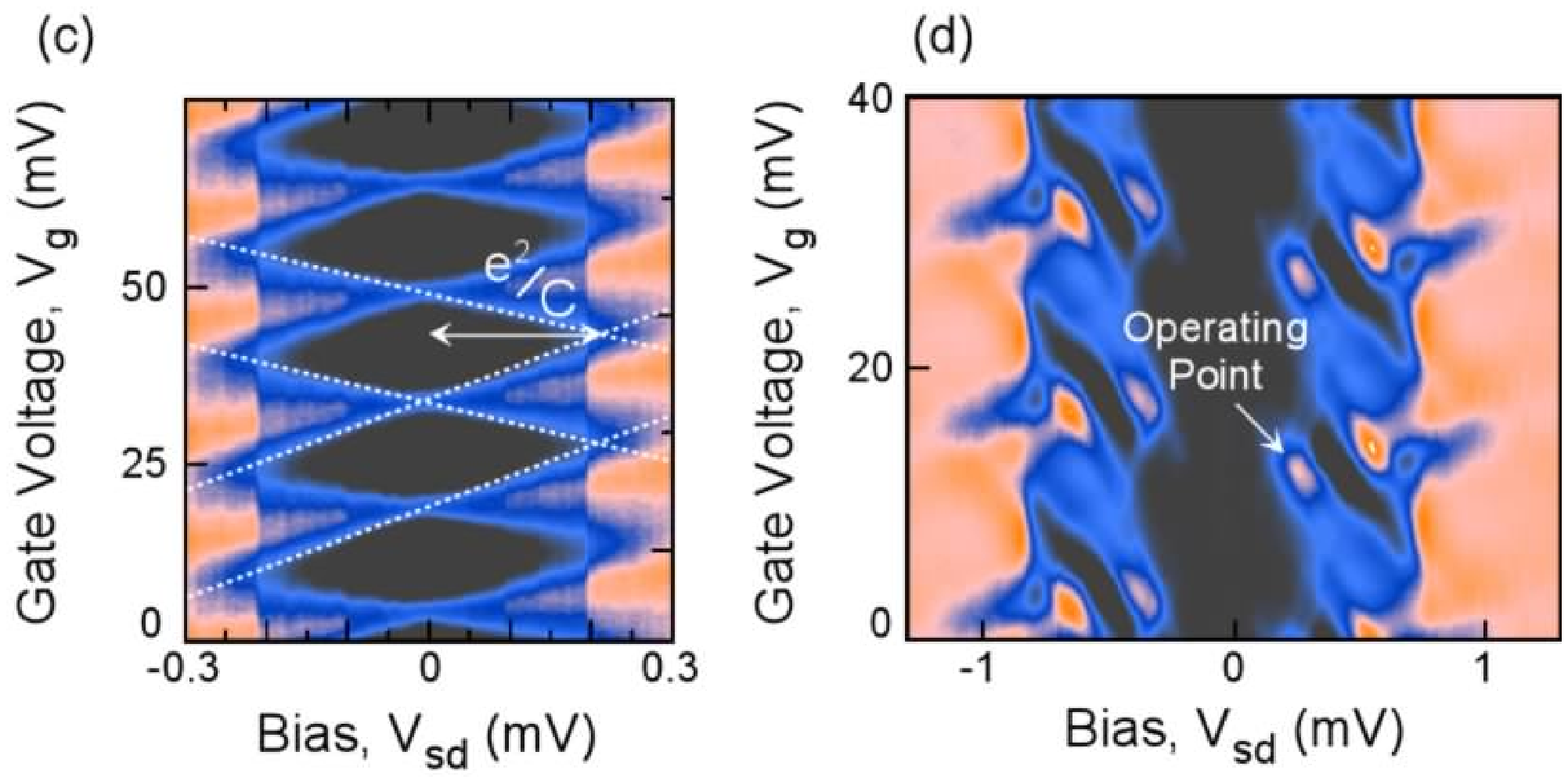}\\
\includegraphics[width=8.5cm]{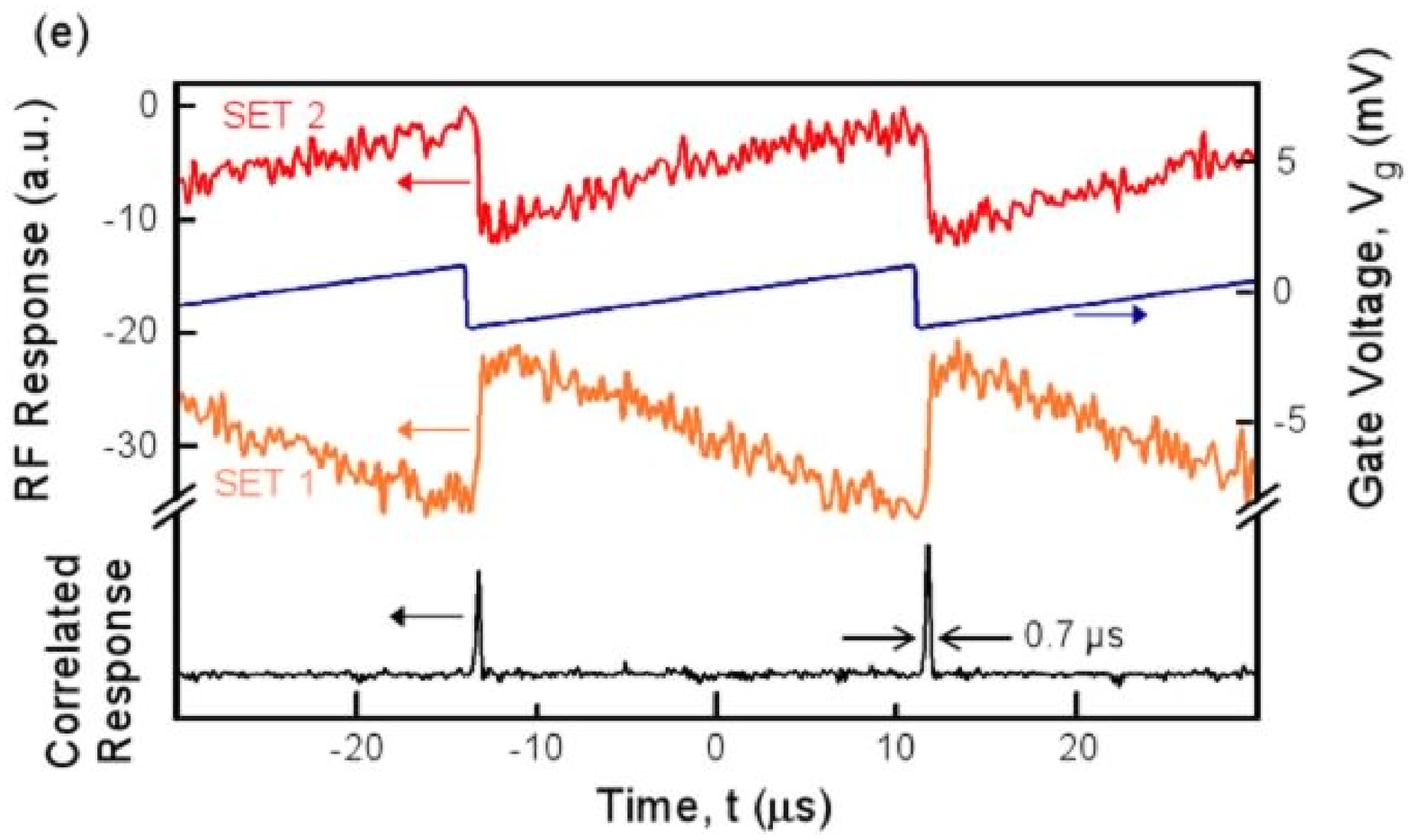}
\caption{(a) Coulomb blockade oscillations in SET conductance on a
control device as in Fig. 3(b), but with no phosphorus implants.
(b) Reflected RF power as a function of frequency on a similar
device when the two SETs are biased at conductance maxima (red)
and minima (black). (c, d) RF bias spectroscopy of SET 2 on a
control device in the normal state (c), and the superconducting
state (d), measured using a carrier frequency of 357 MHz. Each
plot contains 250,000 data points, accumulated at a rate of 0.3 ms
per datum. (e) Reflected RF power for two SETs on a device with a
modified gate arrangement, together with their cross-correlated
output (black), showing sub-${\rm \mu}$s response to a sawtooth
potential (blue) applied to a control gate to simulate charge
transfer.}
\end{figure}\label{fig4}

When applied to the device of Fig. 3(b) this RF measurement system
enables independent operation of the two SETs on microsecond
timescales. Fig. 4(a) shows typical Coulomb blockade oscillations
in the d.c. source-drain conductance of SET 1 on a control device
with no implants as a function of $S$-gate bias, while Fig. 4(b)
shows the reflected RF power as a function of frequency under two
different bias conditions. The red trace corresponds to the
situation when {\it both} SETs are biased to a conductance peak,
while the black trace corresponds to both SETs at conductance
minima. SET 1 is impedance-matched to a 50 $\Omega$ transmission
line at 322 MHz while SET 2 is matched at 357 MHz. By supplying
and monitoring two independent RF carrier signals we then use
wavelength division multiplexing \cite{a31} to independently
monitor each SET, providing nanosecond time resolution for fast
read-out.

The devices are further characterised using bias spectroscopy
measurements in which a d.c. source-drain voltage $V_{sd}$ is
applied to an SET and its response monitored. Fig. 4 shows bias
spectroscopy for a control device SET, operated at radio
frequencies in the normal ($B$ = 2T) and superconducting ($B$ = 0)
states, respectively. In the normal state (Fig. 4(c)) we see a
characteristic ``Coulomb diamond'' spectrum, from which a charging
energy of e${\rm ^{2}}$/C = 0.2 meV may be obtained. In the
superconducting state (Fig. 4(d)) the data is more complex and
exhibits features due to resonant Josephson effects \cite{a34}.

From measurements on all-aluminium devices \cite{a31} we find that
optimal sensitivity to charge motion is obtained by operating the
two SETs in the superconducting state and biasing each to a
Josephson quasiparticle peak, as shown in Fig. 4(d). In this way
we have attained sensitivities with twin SETs of 4.4 ${\rm
\mu}$e/$\surd$Hz and 7.5 ${\rm \mu}$e/$\surd$Hz respectively
\cite{a31}, close to the quantum limited sensitivity \cite{a35}.
Fig. 4(e) shows RF data obtained on an all-aluminium device when a
control gate is ramped with a sawtooth potential to {\it simulate}
periodic charge transfer. The gate voltage amplitude corresponds
to an induced charge of ${\rm \sim}$0.1$e$ on each SET island. In
addition to the {\it single-shot} (non-averaged) RF output from
each SET, Fig. 4(e) shows the cross-correlated signal obtained by
simply multiplying the time derivatives of the two outputs. Much
of the noise detected by the individual SETs is rejected by the
correlation and the combined system shows a response time below 1
${\rm \mu}$s, as required for charge qubit readout.

\section{Charge qubit operation and coupling}
We now discuss possible one- and two-qubit operations which will
be accessible using the devices shown in Fig. 3(b) when configured
with just two activated phosphorus donors. There exist two choices
for the basis of logical qubit states corresponding to the lowest
two states being localised or de-localised. Since SET readout is
most easily carried out for localised states, we focus here on the
configuration with non-zero $S$-gate biases ($V_{S} {\rm \neq 0}$
in Fig. 5(a)), which serves to localize the electron into the
qubit states $ {\rm {\mid}0{\rangle} = {\mid}L{\rangle}} $ and
${\rm {\mid}1{\rangle} = {\mid}R{\rangle}}$. Calculations
\cite{a13} show that for equal and opposite applied voltages on
the $S$-gates of order 0.1 V the fidelity of qubit definition is
optimal, with mixing of higher states less than 10${\rm ^{-4}}$.
We discuss later the alternative delocalised basis choice ${\rm
{\mid}0{\rangle} = ({\mid}L{\rangle} +
{\mid}R{\rangle})/{\surd}2}$ and ${\rm {\mid}1{\rangle} =
({\mid}L{\rangle} - {\mid}R{\rangle})/{\surd}2}$ in the context of
reducing the effects of decoherence.

To perform a single qubit ${\rm \pi}$/2 rotation in the localized
basis the $S$-gate biases are adjusted to zero, to symmetrize the
potential, and the $B$-gate bias made negative to raise the
barrier and slow the coherent oscillations ($V_{S}$ = 0 in Fig.
5(a)). Typically, such operations will require gate bias precision
down to the mV level. The time for a ${\rm \pi}$/2 rotation will
depend primarily on the donor spacing and chosen $B$-gate bias.
For a typical bias $V_{B}$ ${\rm \sim}$0.5 V we calculate
\cite{a13} rotation times longer than 50 ps, which are accessible
using fast pulse generation technology.

Before the qubit is operational it must be pre-initialised after
fabrication to remove one of the electrons from the ${\rm P-P}$
system. The pre-initialisation process is carried out using the
$S$ and $B$ gates: the electron in the right-hand donor well is
ionized to the continuum by a negative bias ${\rm \sim}$-0.4V on
the right-hand $S$-gate \cite{a13}, while the electron in the left
donor well is partially screened by grounding the $B$-gate and
left-hand $S$-gate. Once the qubit has been pre-initialised, the
two SET outputs can be calibrated for the ${\rm {\mid}}$L${\rm
{\rangle}}$ and ${\rm {\mid}}$R${\rm {\rangle}}$ states.
Thereafter, initialisation of the charge qubit into the ground
state ${\rm {\mid}}$0${\rm {\rangle}}$  = ${\rm {\mid}}$L${\rm
{\rangle}}$ is effected by simply biasing the $S$-gates and
observing the SET outputs.

In the SET readout process we wish to make a projective
measurement of a general state ${{\rm {\mid}\Psi{\rangle}} =
c_{0}{\rm {\mid}0{\rangle}}}$ + ${c_{1}{\rm {\mid}1{\rangle}}}$,
resulting from a sequence of gate operations with the SETs
blockaded so that no current flows \cite{a34,a35}. As a result of
measurement we obtain ${\rm {\mid}0{\rangle}}$ or ${\rm
{\mid}1{\rangle}}$ with probabilities ${\rm {\mid}}c_{0}{\rm
{\mid}^{2}}$ and ${\rm {\mid}}c_{1}{\rm {\mid}^{2}}$ respectively.
When voltages are applied to the SET bias gates (see Fig. 3(b))
tuning each SET to a conductance peak, the current flow through
the device decoheres the charge qubit strongly, causing a rapid
transition to a statistical mixture of the localised eigenstates.
The SETs will then give distinguishable output signals, determined
during calibration, corresponding to the system having collapsed
into the left or right state.

\begin{figure}
\centering
\includegraphics[width=8.5cm]{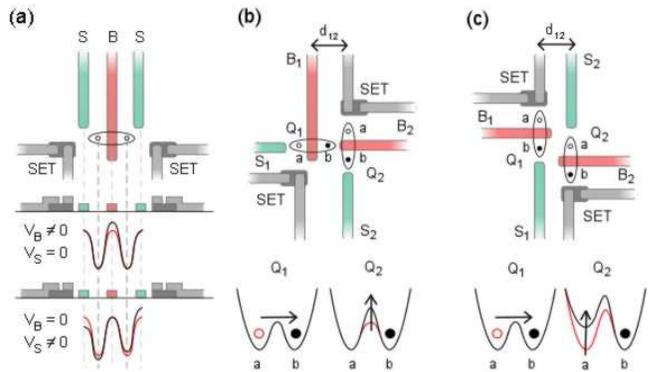}
\caption{(a) Operation of a single charge-qubit showing barrier
($B$-gate) and symmetry ($S$-gates) control, with twin-SET for
correlated charge-position readout. (b, c) Qubit coupling schemes
based on the Coulomb interaction. (b) ``CNOT'' configuration,
where the control qubit $Q_{1}$ alters the effective barrier
height as seen by the target qubit $Q_{2}$. (c) ``CPHASE''
configuration, where the control qubit $Q_{1}$ alters the
potential symmetry at the target qubit $Q_{2}$.}
\end{figure}\label{fig5}

Two possible qubit coupling schemes are shown in Figs. 5(b) and
5(c), leading to different coupling dynamics. For simplicity we
have included only one SET readout device per qubit in the
schematics. In the first ``CNOT'' arrangement, the horizontal
qubit $Q_{1}$ acts on the effective barrier height of the vertical
qubit $Q_{2}$, and the coupling is primarily
${\Gamma}_{zx}{\sigma}_{z}{\rm ^{(1)}}{\sigma}_{x}{\rm ^{(2)}}$.
In the second ``CPHASE'' arrangement the effective coupling is
${\Gamma}_{zz}{\sigma}_{z}{\rm ^{(1)}}{\sigma}_{z}{\rm ^{(2)}}$.
Precise determination of the coupled qubit dynamics in the
presence of the full gate structure is beyond the scope of this
paper, however, we have performed a semi-classical calculation to
obtain an order of magnitude estimate. We calculate the effect on
$Q_{2}$ of moving a charge of 1.0 $e$ between the $a$ and $b$
positions of $Q_{1}$ (see Fig. 5(b)) with $a$ and $b$ chosen to be
60 nm and 30 nm respectively from $Q_{2}$, and obtain ${\Gamma
\approx}$ ${\rm 10^{-4} - 10^{-5}}$ meV corresponding to coupled
qubit operation times of 0.1 - 1 ns.

\section{Discussion}
Given a capability to implant controlled numbers of donors into
silicon devices with nanoscale gates there do not appear to be any
fundamental limits posed by fabrication technologies for scale-up
of this architecture to many qubits. For example, long linear
arrays of charge qubits can be envisaged \cite{a13} with
``CPHASE'' coupling as in Fig. 5(c). Such devices would require a
focused phosphorus ion beam combined with an EBL-defined resist
aperture mask to position each dopant at the appropriate array
site in a step-and-repeat process.

Some device-related issues still require further experimental work
before Si:P qubit operation can be demonstrated. For example, it
will be necessary to minimise defects at the Si/SiO${\rm _{2}}$
interface which could trap the sole electron in the qubit.
Interface trap densities below ${\rm 10^{9} cm^{-2}}$ have been
reported \cite{a36} that correspond to trap spacings in excess of
300 nm, sufficient for qubit operation, however, such interfaces
require high levels of purity during oxide growth. Also, to ensure
that only the intended phosphorus ions enter the device during
implantation, the detector dark currents must be further reduced.
We are currently incorporating $p$- and $n$-doped wells below the
detector electrodes to create a $p$-$i$-$n$ structure for this
purpose.

The key factor determining the limit to scale-up will be
decoherence of the qubits due to environmental coupling.
Successful operation of quantum devices is contingent on coherence
times remaining longer than the time required for arbitrary
rotations. The primary sources of decoherence for the charge qubit
are expected to include phonons, Johnson noise from the gates and
charge noise from the material environment. A recent calculation
of LA phonon induced decoherence on the ${\rm P-P^{+}}$ system
\cite{a37} concluded that for donor separations of 25 nm and
greater, $\tau_{phonon}$ is of order ${\rm \mu}$s, well above the
gate operation times. Similarly, Johnson noise due to fluctuations
on the $S$ and $B$ gates is also calculated \cite{a13,a38} to give
decoherence on timescales ${\rm \sim}$1 ${\rm \mu}$s.

The 1/$f$ noise resulting from charge fluctuations in the
surrounding environment is believed to be the limiting factor for
all charge-based qubits \cite{a39}. In particular, individual
charge traps can produce sudden and large changes in the noise
signal at random times (random telegraph signals). Ensemble
coherence times of  ${\tau {\rm _{2}^{*}}}$ ${\rm \sim}$ 1 ns have
been observed in GaAs quantum dots \cite{a40} suggesting a lower
bound for the single particle coherence time of ${\tau {\rm
_{2}}}$ ${>}$ 1 ns. In comparison with the buried Si:P charge
qubit, these lithographically-defined GaAs devices were quite
large - approximately 0.02 ${\rm \mu}$m${\rm ^{2}}$ in area with
around 25 electrons per dot. We therefore expect the quantum
coherence time for the Si:P charge qubit to be long enough to
allow coherent control of devices such as in Fig. 3.

The use of high quality materials with low trap densities should
also extend the coherence time, while refocusing pulse techniques
can minimise the effects of decoherence \cite{a41}. Furthermore,
${\tau {\rm _{2}}}$ may be increased by defining the logical
states in terms of de-localised symmetric and anti-symmetric
wavefunctions. Since the delocalised states have a very similar
charge distribution they will be less vulnerable to environmental
charge fluctuations \cite{a5}, although charge readout will be
more difficult in this case.

Prior to coherent gate operations two key experiments will be
undertaken. Controlled tunneling between clusters of phosphorus
donors must first be demonstrated with SET readout to ascertain
the signal level corresponding to single electron transfer.
Microwave spectroscopy \cite{a30} on a ${\rm P-P^{+}}$ device may
then be used to map out the qubit energy levels and determine
${\tau {\rm _{1}}}$ and ${\tau {\rm _{2}^{*}}}$ for the system.

In summary, a top-down process for constructing small-scale Si:P
quantum computer devices has been developed that is capable of
accurately locating single phosphorus atoms in an $i$-Si substrate
below custom-configured control gates and SET readout devices. Ion
detection electrodes, integrated into the device structure, make
this single atom doping possible and open the way for a new class
of implanted devices at the single donor level. We have outlined a
scheme for operation of these Si:P devices as charge qubits,
including coupling architectures suitable for scalable computing,
and have discussed in detail the operation of twin RF SETs for
state readout with charge noise rejection. With these fabrication
and measurement technologies in place, a path is open for the
realisation of quantum gate operations in silicon.

\section*{Acknowledgments}
We thank R. Brenner, D. P. George, H. S. Goan, F. Green, A.
Greentree, W. K. Hew, T. Hopf, J. C. McCallum, L. D. Macks, M.
Mitic, T. C. Ralph, L. Sim, P. Spizzirri and R. P. Starrett for
experimental and theoretical contributions. This work was
supported by the Australian Research Council, the Australian
Government, the U.S. National Security Agency (NSA), the Advanced
Research and Development Activity (ARDA) and the Army Research
Office (ARO) under contract number DAAD19-01-1-0653.

\pagebreak


\begin{thebibliography}{99}
%\bibitem{a1} Steane A {\it Rep. Prog. Phys.} {\bf 61} 117 (1998)

\bibitem{a1} For a review see: Y. Makhlin, G. Sch\"on, and A.
Shnirman, Quantum-state engineering with Josephson-junction
devices {\it Rev. Mod. Phys.} {\bf 73}, 357 (2001).

\bibitem{a2} Y. Nakamura, Yu A. Pashkin, and J. S. Tsai, Coherent
control of macroscopic quantum states in a single Cooper-pair box
{\it Nature} {\bf 398}, 786 (1999).

\bibitem{a3} J. E. Mooij, T. P. Orlando, L. Levitov, L. Tian, C. H. van der
Wal, and S. Lloyd, Josephson persistent-current qubit {\it
Science} {\bf 285}, 1036 (1999).

\bibitem{a4} J. R. Friedman, V. Patel, W. Chen, S. K. Tolpygo, and
J. E. Lukens, Quantum superposition of distinct macroscopic states
{\it Nature} {\bf 406}, 43 (2000).

\bibitem{a5} D. Vion, A. Aassime, A. Cottet, P. Joyez, H. Pothier,
C. Urbina, D. Esteve, and M. H. Devoret, Manipulating the quantum
state of an electrical circuit {\it Science} {\bf 296}, 886
(2002).

\bibitem{a6} J. M. Martinis, S. Nam, J. Aumentado, and C. Urbina, Rabi
oscillations in a large Josephson-junction qubit {\it Phys. Rev.
Lett.} {\bf 89}, 117901 (2002).

\bibitem{a7} Y. A. Pashkin, T. Yamamoto, O. Astafiev, Y. Nakamura,
D. V. Averin, and J. S Tsai, Quantum oscillations in two coupled
charge qubits {\it Nature} {\bf 421}, 823 (2003).

\bibitem{a8} B. E. Kane, A silicon-based nuclear spin quantum computer
{\it Nature} {\bf 393}, 133 (1998).

\bibitem{a9} R. Vrijen, E. Yablonovitch, K. Wang, H. W. Jiang,
A. Balandin, V. Roychowdhury, T. Mor, and D. DiVincenzo,
Electron-spin-resonance transistors for quantum computing in
silicon-germanium heterostructures {\it Phys. Rev. A} {\bf 62},
012306 (2000).

\bibitem{a10} J. Wrachtrup et al., unpublished.

\bibitem{a11} R. J. Schoelkopf, P. Wahlgren, A. A. Kozhevnikov, P.
Delsing, and D. E. Prober, The Radio-Frequency Single-Electron
Transistor (RF-SET): A Fast and Ultrasensitive Electrometer {\it
Science} {\bf 280}, 1238 (1998).

\bibitem{a12} A. Aassime, G. Johansson, G. Wendin, R. J.
Schoelkopf, and P. Delsing, Radio-frequency single-electron
transistor as readout device for qubits: charge sensitivity and
backaction {\it Phys. Rev. Lett.} {\bf 86}, 3376 (2001).

\bibitem{a13} L. C. L. Hollenberg, A. S. Dzurak, C. Wellard, A. R.
Hamilton, D. J. Reilly, G. J. Milburn, and R. G. Clark,
Charge-based quantum computing using single donors in
semiconductors, {\it arXiv:cond-mat/0306235} (2003).

\bibitem{a14} A. Barenco, D. Deutsch, A. Ekert, and R. Jozsa, Conditional
quantum dynamics and logic gates {\it Phys. Rev. Lett.} {\bf 74},
4083 (1995).

\bibitem{a15} R. Landauer, Minimal energy requirements in communication
{\it Science} {\bf 272}, 1914 (1996).

\bibitem{a16} A. Fowler, C. Wellard, L. C. L. Hollenberg, Error rate of the
Kane quantum computer controlled-NOT gate in the presence of
dephasing {\it Phys. Rev. A} {\bf 67}, 012301 (2003).

\bibitem{a17} J. L. O'Brien, S. R. Schofield, M. Y. Simmons, R. G. Clark,
A. S. Dzurak, N. J. Curson, B. E. Kane, N. S. McAlpine, M. E.
Hawley, and G. W. Brown, Towards the fabrication of phosphorous
qubits for a silicon quantum computer {\it Phys. Rev. B} {\bf 64},
161401 (2001).

\bibitem{a18} S. R. Schofield, N. J. Curson, M. Y. Simmons, L. Oberbeck,
T. Hallam, F. J. Ruess, and R. G. Clark, Atomically precise placement of single dopants in
Si, unpublished.

\bibitem{a18a} T. M. B\"{u}hler, R. P. McKinnon, N. E. Lumpkin, R.
Brenner, D. J. Reilly, L. D. Macks, A. R. Hamilton, A. S. Dzurak,
and R. G. Clark, A self-aligned fabrication process for silicon
quantum computer devices {\it Nanotechnology} {\bf 13}, 686
(2002); R. P. McKinnon, F. E. Stanley, N. E. Lumpkin, E. Gauja, L.
D. Macks, M. Mitic, V. Chan, K. Peceros, T. M. B\"{u}hler, A. S.
Dzurak, R. G. Clark, C. Yang, D. N. Jamieson, and S. Prawer,
Nanofabrication processes for single-ion implantation of silicon
quantum computer devices {\it Smart. Mater. Struct.} {\bf 11}, 735
(2002).

\bibitem{a19} C. I. Pakes, D. P. George, D. N. Jamieson, C. Yang,
A. S. Dzurak, E. Gauja, and R. G. Clark, Technology computer aided
design modeling of single atom doping for fabrication of buried
nanostructures {\it Nanotechnology} {\bf 14}, 157 (2003).

\bibitem{a20} J. F. Ziegler, J. P. Biersack, and U. Littmark, SRIM - The
Stopping and Range of Ions in Solids {\it Pergamon Press, New
York} (1985).

\bibitem{a21} TCAD - Technology Computer Aided Design, Integrated Systems
Engineering AG, Zurich.

\bibitem{a22} C. Yang, D. N. Jamieson, S. M. Hearne, C. I. Pakes,
B. Rout, E. Gauja, A. S. Dzurak, and R. G. Clark, Ion-beam-induced
charge characterisation of particle detectors {\it Nucl. Instrum.
Meth. B} {\bf 190}, 212 (2002).

\bibitem{a23} G. Bertolini, and A. Coche, Semiconductor Detectors,
{\it North-Holland, Amsterdam} (1968).

\bibitem{a24} K. C. Hsieh, Observation of the pulse height defect for helium
ions of energies <30 keV in silicon {\it Nucl. Instr. And Meth.}
{\bf 138}, 677 (1976); C. C. Curtis, and K. C. Hsieh, Field
ionization device used for measuring the pulse height defect of
He+ in silicon {\it Rev. Sci. Instr.} {\bf 48}, 252 (1977).

\bibitem{a25} V. E. Borisenko, and P. J. Hesketh, Rapid Thermal Processing of
Semiconductors {\it Plenum Press, New York} (1997).

\bibitem{a26} Intrinsic diffusivity calculated for P with values taken from:
P. M. Fahey, P. B. Griffin, and J. D. Plummer, Defects and dopant
diffusion in silicon {\it Rev. Mod. Phys.} {\bf 61}, 289 (1989).

\bibitem{a27} We have confirmed the reliable low-temperature conductivity at
4.2 K of these narrow Ti/Au gates using separate structures.

\bibitem{a28} T. A. Fulton, and G. J. Dolan, Observation of single-electron
charging effects in small tunnel junctions {\it Phys. Rev. Lett.}
{\bf 59}, 109 (1987).

\bibitem{a29} T. M. Buehler, D. J. Reilly, R. Brenner, A. R, Hamilton,
A. S. Dzurak and R. G. Clark, Correlated charge detection for
readout of a solid-state quantum computer {\it Appl. Phys. Lett.}
{\bf 82}, 577 (2003).

\bibitem{a30} K. W. Lehnert, K. Bladh, L. F. Spietz, D. Gunnarsson,
D. I. Schuster, P. Delesing, and R. J. Schoelkopf, Measurement of
the excited-state lifetime of a microelectronic circuit {\it Phys.
Rev. Lett.} {\bf 90}, 027002 (2003).

\bibitem{a31} T. M. Buehler, D. J. Reilly, R. P. Starrett, A. R. Hamilton,
A. S. Dzurak, and R. G. Clark, Development and operation of a
radio frequency twin-SET for solid-state qubit readout {\it
arXiv:cond-mat/0302085} (2003).

\bibitem{a32} T. M. Buehler, D. J. Reilly, R. P. Starrett, A. D. Greentree,
A. R. Hamilton, A. S. Dzurak, and R. G. Clark, Efficient readout
with the radio frequency single electron transistor in the
presence of charge noise {\it arXiv:cond-mat/0304384} (2003).

\bibitem{a33} C. I. Pakes, V. Conrad, J. C. Ang, F. Green,
A. S. Dzurak, L. C. L. Hollenberg, D. N. Jamieson, and R. G.
Clark, Modeling SET-based readout in the Kane solid-state quantum
computer {\it Nanotechnology} {\bf 14}, 161 (2003).

\bibitem{a34} A. A. Clerk, S. M. Girvin, A. K. Nguyen, and
A. D. Stone, Resonant Cooper-Pair Tunneling: Quantum Noise and
Measurement Characteristics {\it Phys. Rev. Lett.} {\bf 89},
176804 (2002).

\bibitem{a35} M. H. Devoret, and R. J. Schoelkopf, Amplifying quantum signals
with the single-electron transistor {\it Nature} {\bf 406}, 1039
(2000).

\bibitem{a36} N. S. Saks, Measurement of single interface trap cross
sections with charge pumping {\it Appl. Phys. Lett.} {\bf 70},
3380 (1997).

\bibitem{a37} S. Barrett, and G. J. Milburn, Measuring the decoherence rate
in a semiconductor charge qubit {\it arXiv:cond-mat/0302238}
(2003).

\bibitem{a38} C. Wellard, and L. C. L. Hollenberg, Thermal noise in a
solid-state quantum computer {\it J. Phys. D} {\bf 35}, 2499
(2002).

\bibitem{a39} E. Paladino, L. Faoro, G. Falci, and R. Fazio, Decoherence
and 1/f Noise in Josephson Qubits {\it Phys. Rev. Lett.} {\bf 88},
228304 (2002).

\bibitem{a40} T. H. Oosterkamp, T. Fujisawa, W. G. van der Wiel, K. Ishibashi,
R. V. Hijman, S. Tarucha, and L. P. Kouwenhoven, Microwave
spectroscopy of a quantum-dot molecule {\it Nature} {\bf 395}, 873
(1998).

\bibitem{a41} J. M. Martinis, S. Nam, J. Aumentado, K. M. Lang, and
C. Urbina, Decoherence of a superconducting qubit due to bias
noise {\it Phys. Rev. B} {\bf 67}, 094510 (2003).

 \end{thebibliography}
\end{document}